\documentclass[times,twocolumn,final]{elsarticle}

\usepackage{medima}
\usepackage{framed,multirow}

\usepackage{amssymb}
\usepackage{latexsym}

\usepackage{url}
\usepackage{xcolor}
\usepackage{adjustbox}
\usepackage{hyperref}

\usepackage{cite}
\usepackage{amsmath,amssymb,amsfonts}
\usepackage{graphicx}
\usepackage{textcomp}

\usepackage{algorithmicx}               
\usepackage{algorithm} 
\usepackage{listings}
\usepackage{algpseudocode}  

\usepackage{array}
\usepackage{booktabs}
\usepackage{colortbl}
\definecolor{dark-gray}{gray}{0.7} 
\usepackage{multirow}

\usepackage{cuted}
\usepackage{graphicx, caption}
\usepackage{color,soul}

\definecolor{newcolor}{rgb}{.8,.349,.1}

\journal{Medical Image Analysis}

\begin{document}

\verso{Deng \textit{et~al.}}

\begin{frontmatter}

\title{Cross-scale Multi-instance Learning for Pathological Image Diagnosis}%

\author[1]{Ruining \snm{Deng}}
\author[1]{Can \snm{Cui}}
\author[1]{Lucas W. \snm{Remedios}}
\author[1]{Shunxing \snm{Bao}}
\author[2]{R. Michael \snm{Womick}}
\author[3]{Sophie \snm{Chiron}}
\author[3]{Jia \snm{Li}}
\author[3]{Joseph T. \snm{Roland}}
\author[1]{Ken S. \snm{Lau}}
\author[3]{Qi \snm{Liu}}
\author[3,4]{Keith T. \snm{Wilson}}
\author[3]{Yaohong \snm{Wang}}
\author[3,4]{Lori A. \snm{Coburn}}
\author[1,3]{Bennett A. \snm{Landman}}
\author[1]{Yuankai \snm{Huo}\corref{cor1}}

\cortext[cor1]{Corresponding author. Email: yuankai.huo@vanderbilt.edu}

\address[1]{Vanderbilt University, Nashville, TN 37215, USA}
\address[2]{The University of North Carolina at Chapel Hill, Chapel Hill, NC 27514, USA}
\address[3]{Vanderbilt University Medical Center, Nashville, TN 37232, USA}
\address[4]{Veterans Affairs Tennessee Valley Healthcare System, Nashville, TN 37212, USA}

\received{xxxxxx}
\finalform{xxxxxx}
\accepted{xxxxxx}
\availableonline{xxxxxx}
\communicated{xxxxxx}

\begin{abstract}
Analyzing high resolution whole slide images (WSIs) with regard to information across multiple scales poses a significant challenge in digital pathology. Multi-instance learning (MIL) is a common solution for working with high resolution images by classifying bags of objects (\textit{i.e.} sets of smaller image patches). However, such processing is typically performed at a single scale (\textit{e.g.}, 20$\times$ magnification) of WSIs, disregarding the vital inter-scale information that is key to diagnoses by human pathologists. In this study, we propose a novel cross-scale MIL algorithm to explicitly aggregate inter-scale relationships into a single MIL network for pathological image diagnosis. The contribution of this paper is three-fold: (1) A novel cross-scale MIL (CS-MIL) algorithm that integrates the multi-scale information and the inter-scale relationships is proposed; (2) A toy dataset with scale-specific morphological features is created and released to examine and visualize differential cross-scale attention; (3) Superior performance on both in-house and public datasets is demonstrated by our simple cross-scale MIL strategy. The official implementation is publicly available at \url{https://github.com/hrlblab/CS-MIL}.

\end{abstract}


\end{frontmatter}


\section{Introduction}

\begin{figure}[t]
\begin{center}
\includegraphics[width=0.45\textwidth]{{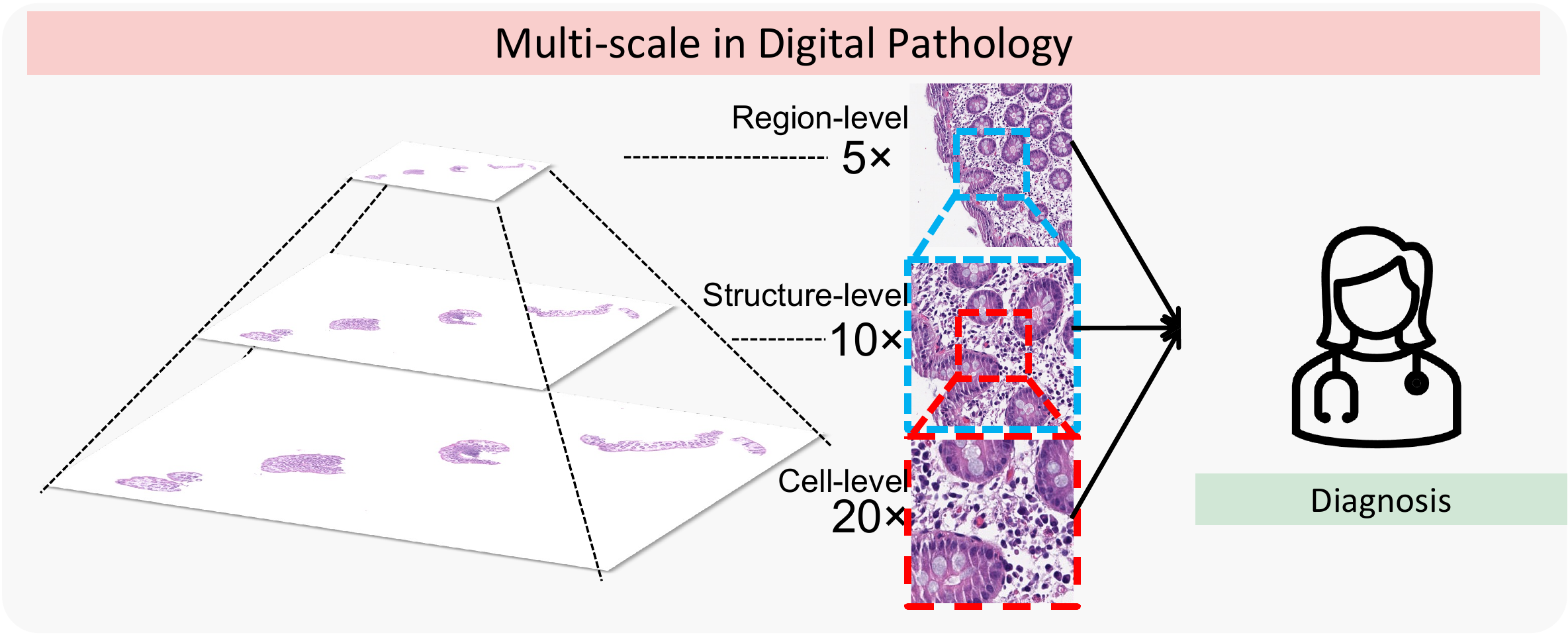}}
\end{center}
\caption{\textbf{Multi-scale awareness.} Given the heterogeneous structural patterns in tissue samples at different resolutions, human pathologists need to carefully examine biopsies at multiple scales across a whole slide image to capture morphological patterns for disease diagnosis. } 
\label{fig1:Problem}
\end{figure}

\begin{figure*}[t]
\begin{center}
\includegraphics[width=1.0\textwidth]{{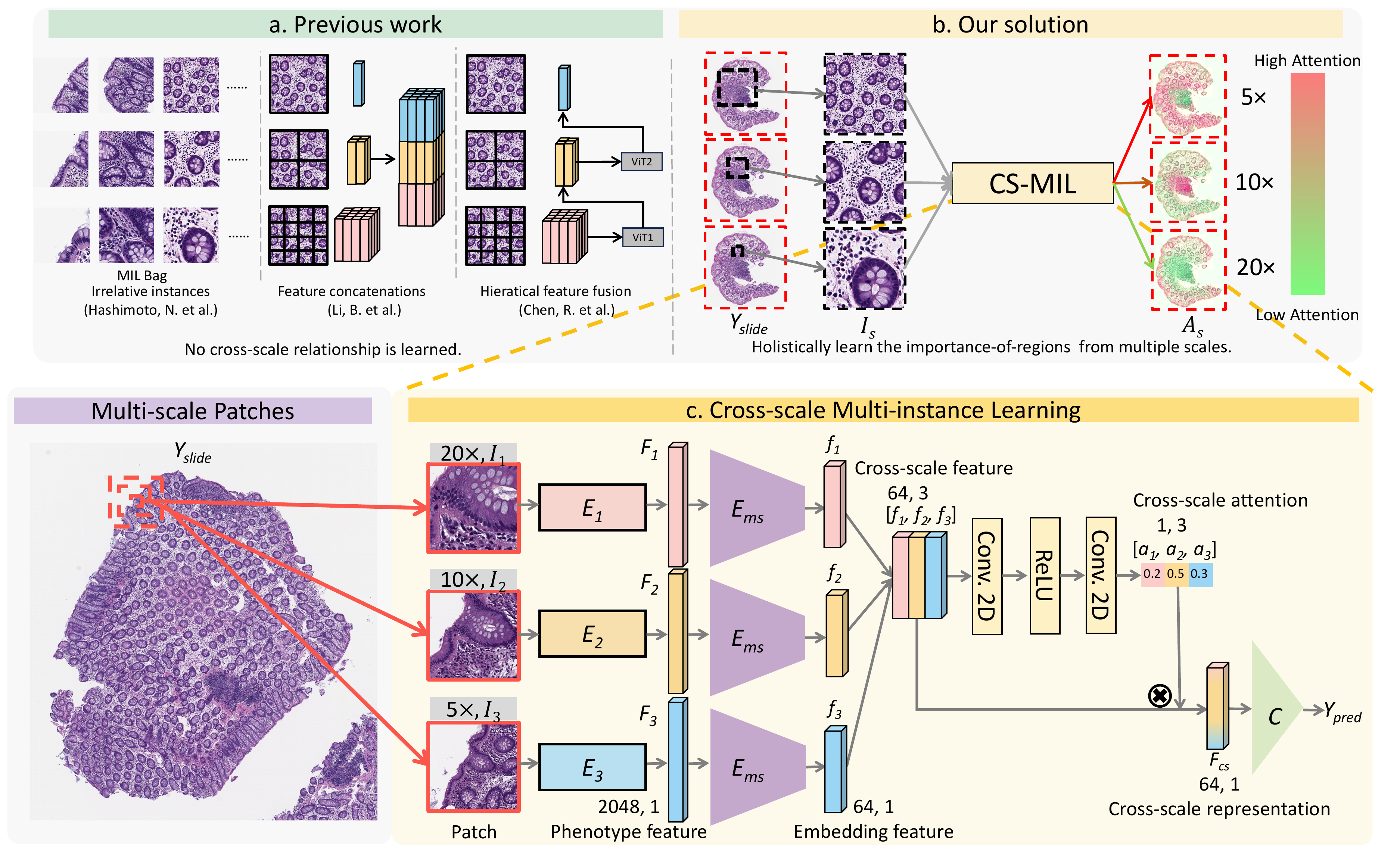}}
\end{center}
\caption{\textbf{Multi-scale MIL designs.} a. Previous work did not take into account the inter-scale relationships across different resolutions. b. Our solution enables the identification of significant regions using cross-scale attention maps, and aggregates the cross-scale features into a cross-scale representation by multiplying the cross-scale attention scores for diagnosing pathological images. c. The cross-scale attention mechanism is employed to merge the cross-scale features with different attention scores. Cross-scale representations from various clusters are concatenated for pathological classification.} 
\label{fig1:Relativework}
\end{figure*}

Pathology is a gold standard to diagnose inflammatory bowel disease (\textit{e.g.}, Crohn's disease)~\citep{gubatan2021artificial,yeshi2020revisiting}. In the current clinical practice, pathologists examine morphological patterns at multiple scales through microscopes~\citep{bejnordi2017diagnostic}, which is a laborious process. With the rapid advancements in whole slide imaging and deep learning techniques, the potential for computer-assisted clinical diagnosis and exploration in digital pathology~\citep{kraszewski2021machine,con2021deep,kiyokawa2022deep,syed2020potential}  is rapidly increasing, making it a promising field of study. However, annotating images pixel- or patch-wise is computationally expensive for a standard supervised deep learning system~\citep{hou2016patch,mousavi2015automated,maksoud2020sos,dimitriou2019deep}. In order to obtain accurate diagnoses from images with weak annotations (\textit{e.g.}, patient-wise diagnosis), multi-instance Learning (MIL) has emerged as a popular weakly supervised learning paradigm for tasks in digital pathology ~\citep{wang2019rmdl,skrede2020deep,chen2021aminn,lu2021data,lu2021ai}. For example, DeepAttnMISL~\citep{yao2020whole} used MIL to cluster image patches into different ``bags" to model and aggregate diverse local features for patient-level diagnosis.

Despite the success, prior efforts--especially ``natural image driven" MIL algorithms--largely overlook the multi-scale (\textit{i.e.} pyramidal) nature of WSIs, which can consist of scales from , thereby allowing pathologists to examine both local and global morphological features~\citep{bejnordi2015multi,gao2016multi,tokunaga2019adaptive}. Recent efforts have been made to mimic human pathological assessments by using multi-scale images in a WSI~\citep{Hashimoto_2020_CVPR,Li_2021_CVPR}. These methods typically extract features independently at each scale and then perform a ``late fusion" step. In this study, we examine the feasibility of introducing interaction between different scales at an earlier stage as an attention-based ``early fusion" paradigm.

Different from the current ``multi-scale" MIL strategy, we propose a novel ``cross-scale" attention mechanism. The key innovation is to introduce an attention-guided MIL scheme to explicitly model inter-scale interactions during feature extraction (Fig.~\ref{fig1:Problem}). The proposed method not only utilizes the morphological features at different scales (with different fields of view), but also learns their inter-scale interactions as an ``early fusion" learning paradigm. Through empirical validation, our cross-scale MIL approach achieves higher Area under the Curve (AUC) scores and Average Precision (AP) scores compared with other multi-scale MIL benchmarks. The study is built upon our earlier work~\citep{deng2022cross}, with a more comprehensive and detailed methodological illustration, a newly released toy dataset, and new validation via a public dataset.

The contribution of this study is three-fold: (1) A novel cross-scale MIL (CS-MIL) algorithm is proposed to explicitly model the inter-scale relationships during feature extraction; (2) A toy dataset with scale-specific morphological features to examine and visualize differential cross-scale attention; (3) Superior performance on both in-house and public datasets is demonstrated by our simple cross-scale MIL strategy. The code has been made publicly available at \url{https://github.com/hrlblab/CS-MIL}.

This work extends our conference paper~\citep{deng2022cross} with the following new efforts and contributions: (1) new experiments are conducted, with three more datasets (one TCGA dataset and two toy datasets) and more recent benchmark methods, for a more rigorous evaluation; (2) a more comprehensive and detailed introduction and illustration of the proposed method are presented in this paper; (3) we release of an end-to-end Docker container for pathological image analysis, facilitating the effective reproduction of the results.

\section{Related Works}
\subsection{Multi-instance Learning in Digital Pathology}
In the realm of clinical digital pathology, disease-related tissue regions may be confined to a relatively small fraction of the entire tissue sample, giving rise to a substantial number of disease-free patches. Pathologists meticulously examine tissues at various magnifications utilizing microscopes to detect the disease-related regions and subsequently scrutinize morphological patterns. Nevertheless, the patch-level annotation of disease-related regions by skilled pathologists is a laborious task that poses challenges in scaling to gigapixel large-scale images. To address this challenge, several recent studies~\citep{hou2016patch,campanella2019clinical,hashimoto2020multi,wang2019rmdl,skrede2020deep,lu2021data,lu2021ai} have demonstrated the promise of weakly supervised technology, multi-instance Learning (MIL) -- a widely used weakly supervised learning paradigm--on patch-level analysis, wherein a patch-based classifier (\textit{e.g.}, patient-wise diagnosis) is trained solely on slide-level labels.

Within the context of MIL, every Whole Slide Image (WSI) is treated as a bag that comprises numerous instances of patches. A WSI bag is marked as disease-relevant if any of its patches (\textit{i.e.}, instances) exhibit disease-related characteristics (\textit{e.g.}, lesions, tumors, abnormal tissues). The classifier refines, extracts, and aggregates patch-level features or scores to anticipate slide-level labels~\citep{Li_2021_CVPR}. Recent MIL based approaches have greatly benefited from using deep neural networks for feature extraction and  aggregation~\citep{ilse2018attention,wang2016deep,oquab2015object}. For example, Yao \textit{et al.},~\citep{yao2020whole} utilized a bag-level approach where image patches were clustered into distinct ``bags" to model and aggregate diverse local features for patient-level diagnosis. In a similar vein, Hou \textit{et al.},~\citep{hou2016patch} proposed a decision fusion model that aggregated patch-level predictions generated by patch-level CNNs. Hashimoto \textit{et al.},~\citep{hashimoto2020multi} proposed a novel CNN-based technique for cancer subtype classification by effectively merging multiple-instance, domain adversarial, and multi-scale learning frameworks at the patch level.

\subsection{Multi-scale in Digital Pathology}

Digital pathology works with pyramidally structured gigapixel images. Different resolutions present different levels of heterogeneous structural patterns on tissue samples. Therefore, pathologists are required to carefully examine biopsies at multiple scales through digital pathology to capture morphological patterns for disease diagnosis~\citep{gordon2020histopathology}. This process is labor-intensive and causes a loss of spatial correlation with sequential zoom-in/zoom-out operations. Using AI models to analyze images at multiple scales not only improves model performance by using scale-aware knowledge but also makes use of inter-scale relationships with spatial consistency learned by the model. 

Previous studies have considered morphological features at multiple scales. Hashimoto \textit{et al.},~\citep{Hashimoto_2020_CVPR} proposed an innovative CNN-based method for cancer subtype classification, effectively integrating multiple-instance, domain adversarial, and multi-scale learning frameworks to combine knowledge from different scales. Li \textit{et al.},~\citep{Li_2021_CVPR} employed a feature concatenation strategy, where high-level features of each region from different scales were merged to incorporate cross-scale morphological patterns obtained from a CNN feature extractor. Barbano \textit{et al.},~\citep{barbano2021unitopatho} proposed a multi-resolution approach for dysplasia grading.

Vision Transformers (ViTs) have emerged as a promising approach for feature learning from large-scale images, owing to their ability to leverage locational attention. Chen \textit{et al.}~\citep{chen2022scaling} recently proposed a novel ViT architecture that exploits the inherent hierarchical structure of WSIs using two levels of self-supervised learning to learn high-resolution image representations. However, none of those methods holistically learn knowledge from multiple scales, that is, regarding inter-scale relationships. To address this limitation, we propose an attention-based ``early fusion" paradigm that offers a promising approach for modeling inter-scale relationships at an early stage.

\section{Methods}
The overall pipeline of the proposed CS-MIL is presented in Fig.~\ref{fig1:Relativework}.c. We propose a novel attention-based ``early fusion" paradigm that aims to capture inter-scale relationships in a holistic manner. First, patches with similar center coordinates but from different scales are jointly tiled from the WSIs. Then, patch-wise phenotype features are extracted using a self-supervised model. Local feature-based clustering is applied to each WSI, which distributes the phenotype patterns into each MIL bag. Next, cross-scale attention-guided MIL is performed to aggregate features across multi-scale and multi-clustered settings. Finally, a cross-scale attention map is generated for human visual examination.

\subsection{Feature embedding and phenotype clustering}

In the MIL community, the majority of histopathological image analysis methods are divided into two stages~\citep{schirris2021deepsmile,dehaene2020self}: (1) the self-supervised feature embedding stage, and (2) the weakly supervised feature-based learning stage. Our approach follows a similar design by utilizing our dataset to train a contrastive-learning model, SimSiam~\citep{chen2021exploring}, as a phenotype encoder ($E_s$) to extract high-level phenotype features ($F_s$) from patches ($I_s$), as shown in equation~\ref{eq:SimSiam}. SimSiam has demonstrated superior feature extraction performance compared to other backbones by maximizing the intra-sample similarity between different image augmentations without any labels. 

\begin{equation}
    F_s = E_s(I_s), s \in (1, S)
\label{eq:SimSiam}
\end{equation}

\noindent where $S$ is the number of the scales on WSIs. Three pretrained encoders ($E_s$) were trained by patches from different scales, respectively. This self-supervised learning stage is crucial for effective feature extraction before the subsequent weakly supervised feature-based learning stage. All of the patches were then embedded into low-dimensional feature vectors for the classification in the second stage.

Inspired by~\citep{yao2020whole}, k-means clustering is used to cluster patches on the patient level based on their self-supervised embeddings at 20$\times$ magnification from the first stage. It is noted that high-level features are more comprehensive than low-resolution thumbnail images in representing phenotypes~\citep{zhu2017wsisa}. The patches were gathered equally from different clusters in each bag and then the bag with better generalization for the MIL model is organized by distinctive phenotype patterns sparsely distributed on WSIs. On the other hand, patches with similar high-level features is aggregated for classification without spatial limitation.

\subsection{Cross-scale attention mechanism}
Our approach builds upon previous work in MIL-related literature by incorporating a cross-scale attention mechanism that captures patterns across scale in whole-slide images (WSIs). Specifically, we utilize an CNN-based encoder to refine patch embeddings from corresponding phenotype clusters. The instance-wise features are then aggregated to achieve patient-wise classification, resulting in superior performance on survival prediction with WSIs. While attention mechanisms have been proposed in previous work to enhance the models' use of patterns across spatial locations in WSIs~\citep{ilse2018attention, lu2021data}, they do not take advantage of patterns across scale in WSIs. Other approaches have aggregated multi-scale features into deep learning models from WSIs~\citep{Hashimoto_2020_CVPR, Li_2021_CVPR}, but have demonstrated limitations in their ability to leverage the interplay between multiple resolutions within the same location.

To address this issue, we propose a novel cross-scale attention mechanism to represent awareness at different scales in the backbone. Firstly, the embedding cross-scale features ($f_s$) from phenotype encoders ($E_s$) are further processed among different scales by a multi-scale encoder ($E_{MS}$) with a siamese Multiple Instance Fully Convolutional Networks (MI-FCN) from DeepAttnMISL~\citep{yao2020whole} in~\ref{eq:MI-FCN}:

\begin{equation}
    f_s = E_{MS}(F_s), s \in (1, S)
\label{eq:MI-FCN}
\end{equation}

\noindent where $S$ is the number of the scales on WSIs. 
All multi-scale encoders ($E_{MS}$) are weight-shared among the different scales.

Next, a cross-scale attention mechanism is applied to consider the importance of each scale in the cross-scale attention within the same location. Cross-scale features ($f_s$) are simultaneously fed into a cross-scale multi-instance learning network (CS-MIL), which contains two fully convolutional layers with kernel size of 1$\times$1, and an ReLU activation function. The output of CS-MIL is the set of cross-scale attention scores ($a_s$) by considering cross-scale features as a whole. This is achieved using Equation \ref{eq:cross-scale2}:

\begin{equation}
a_s = \frac{\exp{\mathbf{W}^\mathrm{T}ReLU(\mathbf{V}f_s^\mathrm{T})}}{\sum_{s=1}^S\exp{\mathbf{W}^\mathrm{T}ReLU(\mathbf{V}f_s^\mathrm{T})}}
\label{eq:cross-scale2}
\end{equation}

\noindent where $\mathbf{W} \in\mathbb{R}^{L \times 1}$ and $\mathbf{V} \in\mathbb{R}^{L \times M}$ are trainable parameters in the CS-MIL, $L$ is the size of the $E_{MS}$ output $f_s$, $M$ is the output channel of the first layer of CS-MIL, $tanh(.)$ is the tangent element-wise non-linear activation function, and $S$ is the number of the scales on WSIs.

The cross-scale attention scores ($a_s$) are then multiplied with cross-scale features, resulting in a fused cross-scale representation (demonstrated in Equation \ref{eq:cross-scale1}):

\begin{equation}
Fcs = \sum_{s=1}^S a_sf_s
\label{eq:cross-scale1}
\end{equation}

Finally, the attention-based instance-level pooling operator ($C$) from \citep{yao2020whole} is deployed to achieve patient-wise classification with cross-scale embedding in~\ref{eq:prediction}, with a bag size of $n$.

\begin{equation}
    Y_{pred} = C(Fcs_1,Fcs_2...,Fcs_n) 
\label{eq:prediction}
\end{equation}

\subsection{Cross-scale attention visualisation}

The cross-scale attention mechanism produces attention scores ($a_s$) for each region ($I_s$) based on cross-scale features ($f_s$) from the CS-MIL. These attention scores reflect the relative importance of phenotype features at different scales for diagnosis when fusing the cross-scale representation ($Fcs$) for final classification ($C$). By filling these scores back to the corresponding location on WSIs, we obtain an attention map ($A_s$) that combines scale and location information. This map provides insights for disease-guided exploration in various contexts, highlighting the versatility and practicality of the cross-scale mechanism.

\section{Experiments}
\subsection{Data}
\textbf{In-house CD dataset:} 50 H\&E-stained Ascending Colon (AC) Diseased biopsies from ~\citep{bao2021cross} were collected from 20 CD patients along with 30 healthy controls for training. The stained tissues were scanned at 20$\times$ magnification. For the pathological diagnosis, the 20 slides from CD patients were scored as normal, quiescent, mild, moderate, or severe, while the remaining tissue slides from healthy controls were scored as normal. 116 AC biopsies were stained and scanned for testing with the same procedure as the above training set. The biopsies were acquired from 72 CD patients who have no overlap with the patients in the training data.

\textbf{TCGA-GBMLGG dataset:} To demonstrate the generalizebility of our proposed architecture, we conduct experiments on a glioma dataset (GBMLGG) obtained from The Cancer Genome Atlas (TCGA). The dataset contains 613 patient samples, of which 330 patients have Isocitrate dehydrogenase (IDH) mutations, while the remaining patients are normal.

\subsection{Experimental setting}
\textbf{In-house CD dataset:} All WSIs from the two datasets were cropped into regions with the size of 4096 $\times$ 4096 pixels to fairly compare the performance between MIL methods and the ViT method in 20$\times$. For 20$\times$ patches, each pixel is equal to 0.5 Microns. Then, 256$\times$256 pixels patches were tiled at three scales (20$\times$, 10$\times$ and 5$\times$) for those regions. Three individual models following the official SimSiam model with a ResNet-50 backbone were trained at three different scales with all of the patches (504,444 256 $\times$ 256 foreground patches).The training was conducted in 200 epochs with a batch size of 128 with the official settings of the SimSiam. 2048-channel embedding vectors were obtained for all patches. Phenotype clustering was performed within the single-scale features at three resolutions using k-means clustering with a class number of 8, and cross-scale features were generated that included all resolutions for each patient. For feature extraction with HIPT~\citep{chen2022scaling}, the official pre-training implementation was used with 1650 4096 $\times$ 4096 regions..

The training dataset was randomly organized into 10 data splits using a ``leave-one-out" strategy, while the testing dataset was divided into 10 splits with balanced numbers accordingly. MIL models were used to collect each bag for every patient, with an equal selection of different phenotype clustering classes marked with a slide-wise label ($Y_{slide}$) from clinicians. The hyper-parameters for training were consistent with those of DeepAttnMISL~\citep{yao2020whole}. The Negative Log-Likelihood Loss~\citep{yao2019negative} was employed to compare the slide-wise prediction ($Y_{pred}$) for the bag with the weakly label in~\ref{eq:loss}. 

\begin{equation}
    L(\theta) = -\log p(Y_{slide}|Y_{pred};\theta)
\label{eq:loss}
\end{equation}

\noindent where $\theta$ represents the model parameters. All the models were updated every four epochs to smoothly converge the loss and trained in 100 epochs in total. 

The optimal model on each data split was selected based on the validation loss, while the mean performance across 10 data splits was used to evaluate the testing results. During the testing stage, 100 image bags were randomly generated, each with a size of 8 to cover most of the patches on each Whole Slide Image (WSI), and the mean value of bag scores was calculated as the final prediction at the slide level. The performance of each model was estimated using Receiver Operating Characteristic (ROC) curves with Area under the Curve (AUC) scores, Precision-Recall (PR) curves with Average Precision (AP) scores, and classification accuracy. All models were trained on an NVIDIA RTX5000 GPU.

\textbf{TCGA-GBMLGG dataset:} Due to computational constraints, a 10$\%$ area was randomly selected from each WSI, resulting in a dataset of 5132 4096 $\times$ 4096 regions. During the pre-training stage, only 15$\%$ of these regions were used (582,666 256 $\times$ 256 foreground patches at three scales from 755 4096 $\times$ 4096 regions) for training the SimSiam model with a ResNet-50 backbone. The official pre-training parameters and hyper-parameters for HIPT were used in the testing stage, since HIPT already included the TCGA-GBMLGG dataset (with 54158 regions) in its pre-training stage. The training, validation, and testing samples were separated at a patient level with a 6:1:3 ratio. In the testing stage, 500 image bags of size 32 were randomly generated for each WSI, and the mean of the bag scores was calculated as the final prediction at the patient level. All models were trained using NVIDIA RTXA6000 GPU.

\begin{table*}[t]
\caption{Classification Performance on two dataset.}
\begin{center}
\begin{adjustbox}{width=0.95\textwidth}
\begin{tabular}{l|cc|ccc|ccc}
\toprule
\multirow{1}{0.8in}{Method} & \multicolumn{2}{c}{Setting} & \multicolumn{3}{c}{CD} &  \multicolumn{3}{c}{TCGA-GBMLGG} \\
\cmidrule(lr){2-3}
\cmidrule(lr){4-6}
\cmidrule(lr){7-9}
& Patch Scale & Clustering & AUC & AP & $p$-value & AUC & AP & $p$-value\\
\midrule
DeepAttnMISL(20$\times$)~\citep{yao2020whole} & Single & 20$\times$ &  0.7775 & 0.6126 & * & 0.7445 & 0.7750 & *\\
DeepAttnMISL(10$\times$)~\citep{yao2020whole} & Single & 10$\times$ & 0.7900 & 0.7113 & * & 0.7345 &  0.7568 & *\\
DeepAttnMISL(5$\times$)~\citep{yao2020whole} & Single & 5$\times$ & 0.8288 & 0.7314 & * & 0.7509 & 0.7957 & *\\
\midrule
Gated Attention~\citep{ilse2018attention} & Multiple & Multiple & 0.8362 & 0.7544 & * & 0.7440 & 0.7606 & *\\
DeepAttnMISL~\citep{yao2020whole} & Multiple & Multiple & 0.8412 & 0.7701 & * & 0.7200 &  0.7520 & *\\
\midrule
MS-DA-MIL~\citep{Hashimoto_2020_CVPR} & Multiple & 20$\times$ & 0.8565 & 0.8090 & * & 0.7562 & 0.8001 & *\\
DS-MIL~\citep{Li_2021_CVPR} & Multiple & 20$\times$ & 0.8386 & 0.7933 & * & 0.7605 & 0.7998 & *\\
DTFD-MIL~\citep{zhang2022dtfd} & Multiple & 20$\times$ & 0.7046 & 0.5380 & * & 0.7215 & 0.7594 & *\\
HIPT~\citep{chen2022scaling} & Multiple & 20$\times$ & 0.7863 & 0.7459 & * & 0.7102 & 0.7430 & *\\
HAG-MIL~\citep{xiong2023diagnose} & Multiple & 20$\times$ & 0.8374 & 0.7459 & * & 0.7610 & 0.8049 & *\\
CS-MIL(Ours) & Multiple & 20$\times$ & \textbf{0.8774} & \textbf{0.8486} & Ref. &\textbf{0.7737} & \textbf{0.8187} & Ref.\\
\bottomrule
\end{tabular}
\end{adjustbox}
\end{center}
\noindent\text{The bootstrapped two-tailed test and the DeLong test is performed with CS-MIL as the reference (“Ref.”) method.} \\
\noindent\text{“*” represents the significant ($p$ $<$ 0.05) differences, while “N.S.” means the difference is not significant.}\\
\label{tab:Classification}
\end{table*}

\begin{figure*}
\begin{center}
\includegraphics[width=0.9\textwidth]{{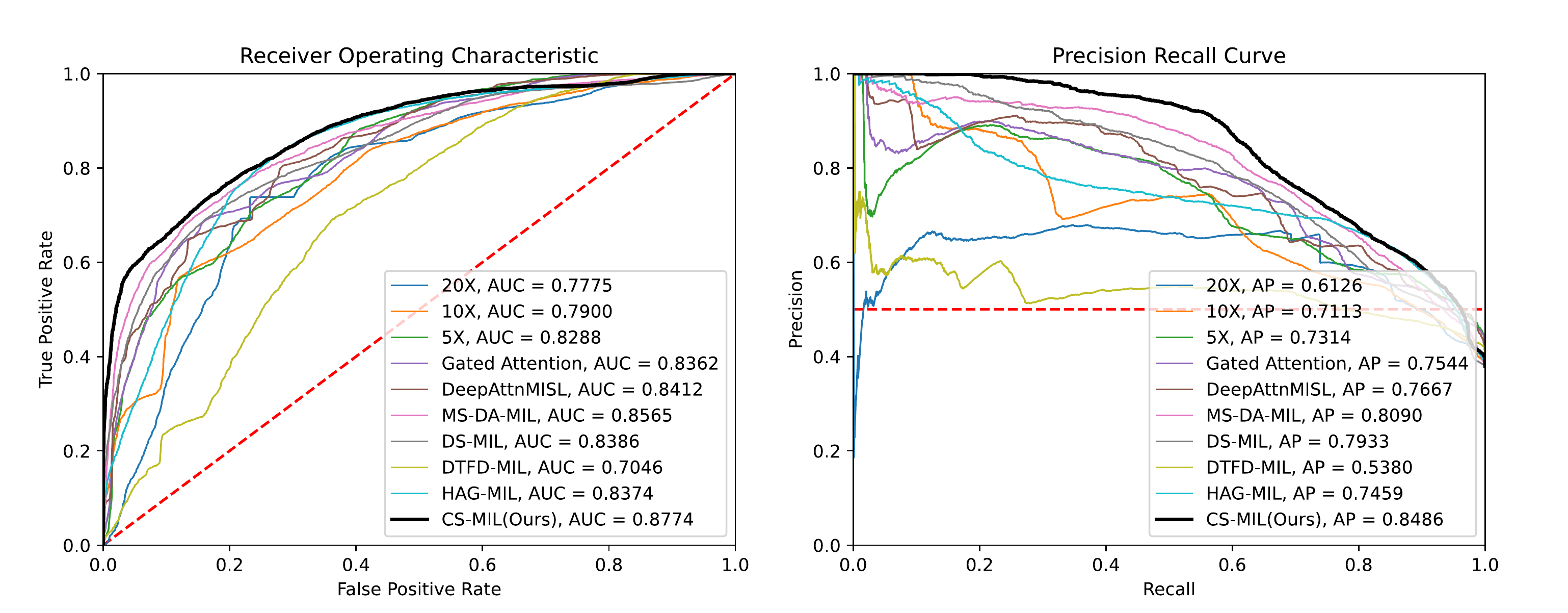}}
\end{center}
\caption{\textbf{ROC curves with AUC scores and PR curves with AP scores.} This figure illustrates the receiver operating characteristic (ROC) curves and precision-recall (PR) curves for both baseline models and the proposed model, along with the corresponding area under the curve (AUC) scores and average precision (AP) scores. The results indicate that the proposed model with cross-scale attention outperformed the baseline models in terms of both metrics.} 
\label{fig3:Results}
\end{figure*}

\section{Results}
\subsection{Empirical Validation}

We implemented three identical single-scale DeepAttnMISL~\citep{yao2020whole} models for patches at corresponding scales. We simultaneously trained the (4) Gated Attention (GA) model ~\citep{ilse2018attention} and (5) DeepAttnMISL model with multi-scale patches, without differentiating scale information. Patches from multiple scales are treated as instances when processing phenotype clustering and patch selection for MIL bags. Furthermore, we adopted multiple multi-scale methods, including (6) a multi-scale feature aggregation (MS-DA-MIL) that jointly adds embedding features from the same location at different scales into each MIL bag~\citep{Hashimoto_2020_CVPR}; (7) a feature concatenation (DS-MIL) at different scales~\citep{Li_2021_CVPR}; (8) A Double-Tier Feature Distillation when aggregating features from multiple scales and multiple locations~\citep{zhang2022dtfd}; (9) a Hierarchical Image Pyramid Transformer (HIPT) with self-supervised learning~\citep{chen2022scaling}; (10) a Hierarchical Attention Guided MIL~\citep{xiong2023diagnose} (HAG-MIL) as well as the proposed method (11) CS-MIL.

We followed above multi-scale aggregation to input phenotype features into the DeepAttnMISL backbone to evaluate the baseline multi-scale MIL models as well as our proposed method. All of the models were trained and validated within the same hyper-parameter setting and data splits.

\begin{figure*}[h]
\begin{center}
\includegraphics[width=0.7\textwidth]{{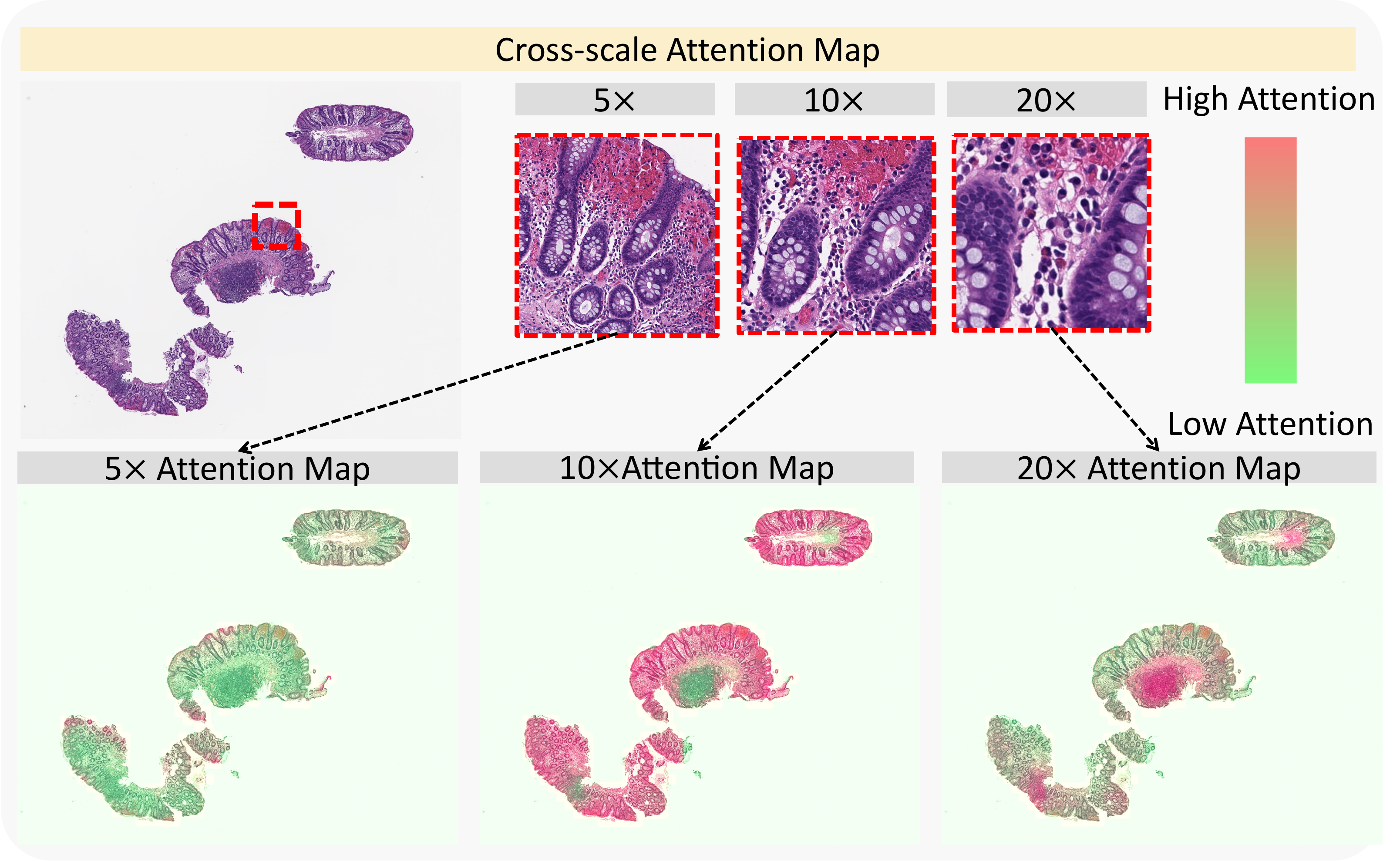}}
\end{center}
\caption{\textbf{Attention Map Visualization.} This figure displays the cross-scale attention maps generated by the proposed model for a CD WSI. The attention map at 20$\times$ highlights the chronic inflammatory infiltrates, whereas the attention map at 10$\times$ focuses on the crypt structures. These regions of interest indicate the distinctive areas for CD diagnosis that are discernible across multiple scales.} 
\label{fig4:AttentionMap}
\end{figure*}

\subsubsection{Classification performance}
Table~\ref{tab:Classification} and Fig.~\ref{fig3:Results} indicate the performance of the classification while directly applying the models on the testing dataset in the CD classification task, without retraining. Table~\ref{tab:Classification} also shows IDH status classification on TCGA-GBMLGG dataset. In general, the proposed CS-MIL achieved better scores in most evaluation metrics, demonstrating the benefits of the cross-scale attention that explores the inter-scale relationship at different scales in MIL. 

\subsubsection{Cross-scale attention visualisation}
Fig.~\ref{fig4:AttentionMap} shows the cross-scale attention maps generated by the proposed CS-MIL on a CD WSI. The proposed CS-MIL presents distinctive importance-of-regions on WSIs at different scales, merging multi-scale and multi-region visualization. As a result, the 20$\times$ attention map highlights the chronic inflammatory infiltrates, while the 10$\times$ attention map focuses on the crypt structures. Those regions of interest interpret the discriminative regions for CD diagnosis across multiple scales.

\subsection{Ablation Studies}
Inspired by ~\citep{yao2020whole} and ~\citep{ilse2018attention}, we explored various attention mechanism designs in MIL, utilizing different activation functions, and evaluated these designs on two datasets. We divided the CS-MIL approach into two strategies, distinguished by whether they share kernel weights during the embedding features learning process across multiple scales. As demonstrated by the classification performance in Table~\ref{tab:Ablation3}, sharing the kernel weights in the CS-MIL strategy, coupled with a ReLU activation function ~\citep{agarap2018deep}, yielded better performance with higher values across various metrics. Overall, the proposed CS-MIL design consistently achieved superior performance compared to other baseline methods. The difference of AUC between each baseline method with our method (``Ref.” in the table) has been statistically evaluated using DeLong test. The significant difference (p $<$ 0.05) are indicated as ``*”, while the non-significant ones are marked as ``N.S.”.

In Table.~\ref{tab:Ablation2}, we further assess the effectiveness of the proposed cross-scale attention mechanism by utilizing a mean-vector and concatenation design with the CS-MIL backbone under a fixed 1:1:1 attention score setting. Additionally, we compare the classification performance when the phenotype embedding is removed from the pipeline, in order to assess the contribution of the projection. As a result, the CS-MIL pipeline equipped with the cross-scale attention mechanism and phenotype clustering during bag generation demonstrated better performance on two empirical datasets.

\begin{table}[t]
\caption{Comparison of different fusion strategies in the multi-scale paradigm.}
\centering
\scriptsize
\setlength{\tabcolsep}{1mm}
\renewcommand\arraystretch{1}
\begin{tabular}{cc|ccc|ccc}
\toprule
\multicolumn{2}{c}{Strategy} & \multicolumn{3}{c}{CD}  & \multicolumn{3}{c}{TCGA-GBMLGG}\\
\cmidrule(lr){1-2}
\cmidrule(lr){3-5}
\cmidrule(lr){6-8}
Layer Kernel & Activation Function & AUC & AP & $p$-value & AUC & AP &$p$-value\\
\midrule
Non-sharing & Tanh & 0.8674 & 0.8436 & * & 0.7562 & 0.8054 & * \\
Non-sharing & ReLU & 0.8686 & 0.8362 & * & 0.7564 & 0.7908 & *  \\
Sharing & Tanh & 0.8652 & 0.8388 & * & 0.7344 & 0.7812 & *  \\
Sharing & ReLU & \textbf{0.8774} & \textbf{0.8486} & Ref. &\textbf{0.7737} & \textbf{0.8187}& Ref.  \\
\bottomrule
\end{tabular}
\noindent\text{“*” represents the significant ($p$ $<$ 0.05) differences,} \\
\noindent\text{while “N.S.” means the difference is not significant.}\\
\label{tab:Ablation3}
\end{table}

\begin{table}[t]
\caption{Learning setting with classification performance on two dataset.}
\centering
\scriptsize
\setlength{\tabcolsep}{1mm}
\renewcommand\arraystretch{1}
\begin{tabular}{cc|ccc|ccc}
\toprule
\multicolumn{2}{c}{Setting} & \multicolumn{3}{c}{CD}  & \multicolumn{3}{c}{TCGA-GBMLGG}\\
\cmidrule(lr){1-2}
\cmidrule(lr){3-5}
\cmidrule(lr){6-8}
Fusion & Clustering & AUC & AP &$p$-value & AUC & AP &$p$-value\\
\midrule
Mean-vector & \checkmark & 0.8147  & 0.7448 & * & 0.7425 & 0.7743 & *\\
Concatenation & \checkmark & 0.8470  &  0.7628 & * & 0.7362 & 0.7664 & *\\
CS-Attention (Naive) & & 0.7692 & 0.7087 & * & 0.7308 & 0.7548 & * \\
CS-Attention (Ours) & \checkmark &\textbf{0.8774} & \textbf{0.8486} & Ref. & \textbf{0.7737} &\textbf{0.8187} & Ref.\\
\bottomrule
\end{tabular}
\noindent\text{“*” represents the significant ($p$ $<$ 0.05) differences, } \\
\noindent\text{while “N.S.” means the difference is not significant.}\\
\label{tab:Ablation2}
\end{table}

\begin{figure}
\begin{center}
\includegraphics[width=0.35\textwidth]{{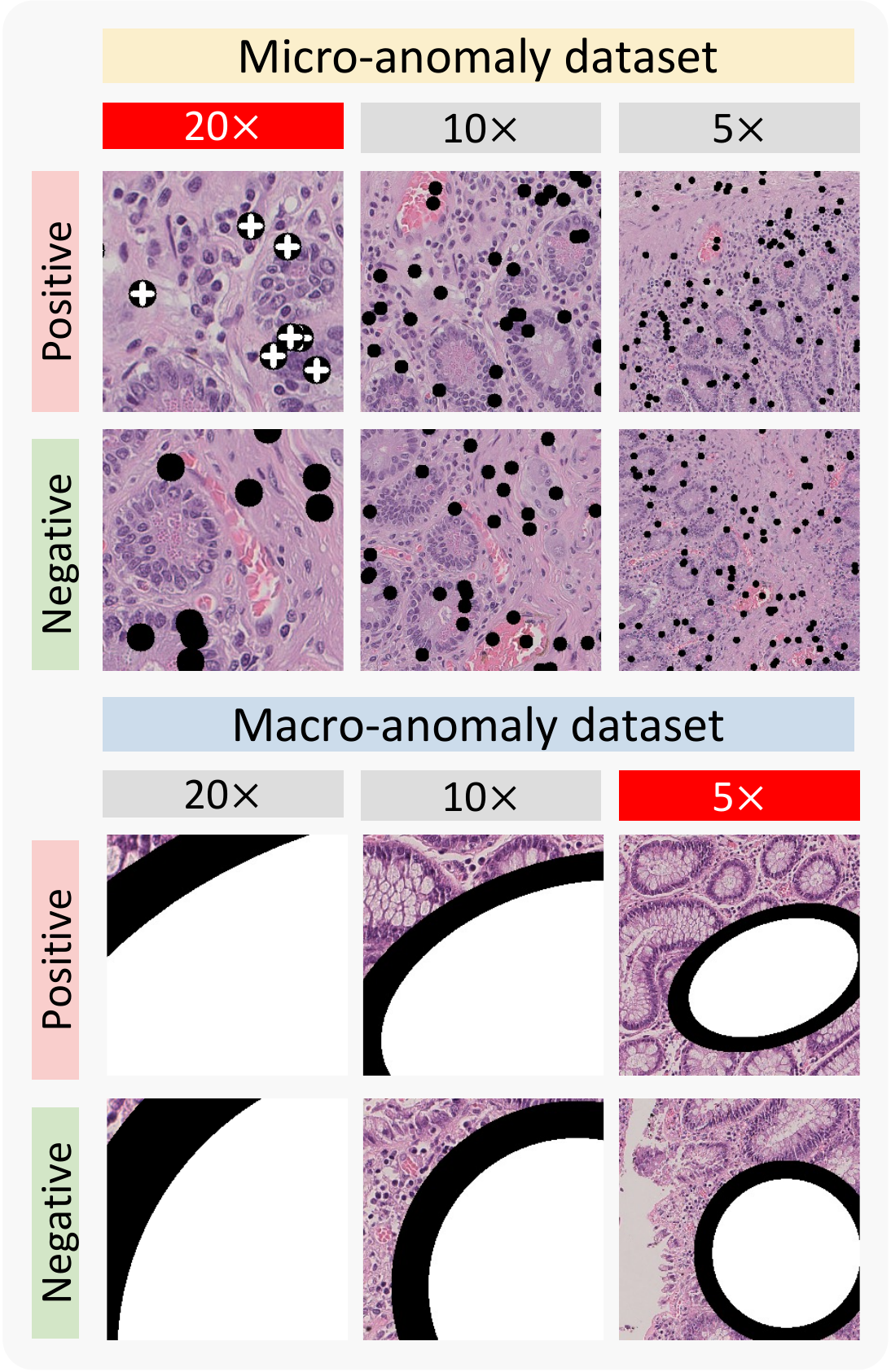}}
\end{center}
\caption{\textbf{Two toy dataset.} This figure demonstrates two toy datasets to evaluate the functionality of the cross-scale attention mechanism. In the micro-anomaly dataset, the white cross pattern is only observed at 20$\times$. In the macro-anomaly dataset, the abnormal shape (ellipse) is easily recognized at 5$\times$.}
\label{fig5:Toydataset}
\end{figure}

\subsection{Simulation}
To assess the effectiveness of the cross-scale attention mechanism, we evaluated CS-MIL using two toy datasets that represent distinct morphological patterns at different scales in digital pathology. These datasets were selected to simulate different scenarios and test the functionality of our approach.

\textbf{Data:} Fig.~\ref{fig5:Toydataset} shows the patches for training in the two datasets (Micro-anomaly dataset and Macro-anomaly dataset). The  micro white crosses pattern only appear on positive patches at 20$\times$ maganification in the micro-anomaly dataset, while the macro anomaly (ellipse) is easily recognized at 5$\times$ with larger visual fields in macro-anomaly dataset. All of the patches are extracted from normal tissue samples in Unitopatho dataset~\citep{9506198}. Two datasets were released to measure the generalization of the cross-scale designs for digital pathology community. The details of two toy datasets are shown in Table.~\ref{tab:toydataset}.

\begin{table}
\centering
\scriptsize
\setlength{\tabcolsep}{1.5mm}
\renewcommand\arraystretch{1}
\caption{The details of two toy datasets.}

\begin{tabular}{l|cccccc}
\toprule
Dataset Id & Training Patches & Validation Patches & Testing Regions & Testing bags\\
\midrule
1 & 5328 & 2772 & 10596 & 4540\\
2 & 2790 & 1548 & 414 & 186\\
\bottomrule
\end{tabular}
\label{tab:toydataset}
\end{table}

\textbf{Approach:}
The CS-MIL utilizes a ResNet-18 backbone to extract features from patches. Our implementation, including hyper-parameters, followed that of DeepAttnMISL~\citep{yao2020whole}. During testing, each patch was randomly captured 10 times into different image bags with a size of 8 to obtain multiple attention scores. The final attention score was calculated by taking the mean value of these scores.

\begin{figure*}[h]
\begin{center}
\includegraphics[width=1.0\textwidth]{{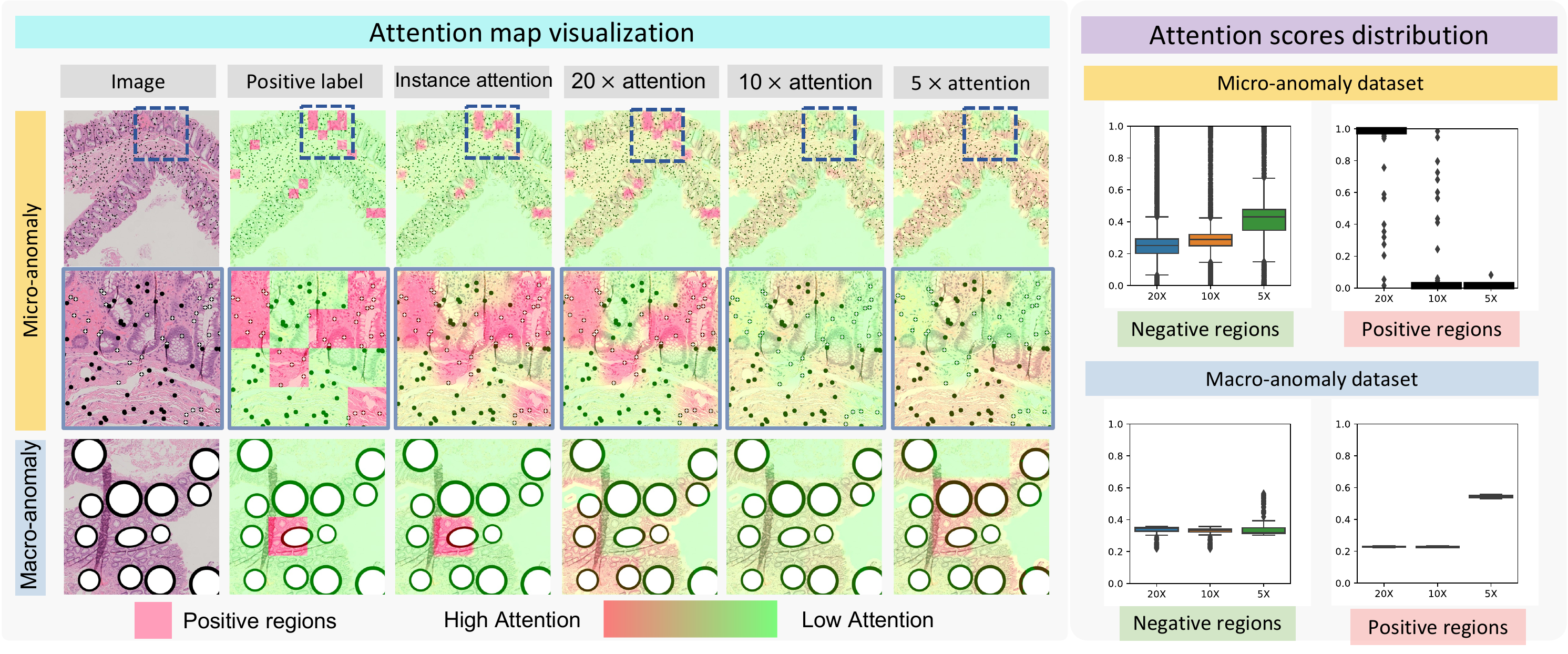}}
\end{center}
\caption{\textbf{Results for toy datasets} This figure exhibits attention maps at both instance level and multiple scales. In the case of the Micro-anomaly dataset, the instance attention generates higher attention scores for the positive regions in their corresponding regions at 20$\times$. Similarly, for the Macro-anomaly dataset, the instance attention identifies ellipses with higher attention scores rather than circles at 5$\times$. Additionally, the box plots in the right panel display the attention score distribution at different scales, confirming the reliability of the cross-scale attention mechanism in generating scores at multiple scales.}
\label{fig5:ToydatasetResults}
\end{figure*}

\textbf{Results:} Table~\ref{tab:ablation1} presents the bag-level classification performance on the two toy datasets. The proposed method differentiates distinctive patterns at different scales in a stable manner. Fig.~\ref{fig5:ToydatasetResults} displays the cross-scale attention maps at the instance level and multiple scales. For the Micro-anomaly dataset, the instance attention successfully highlights positive regions with higher attention scores in corresponding regions at 20$\times$. For the Macro-anomaly dataset, the instance attention locates ellipses instead of circles with higher attention scores at 5$\times$. The box plots in the right panel illustrate the attention score distribution at different scales, attesting to the cross-scale attention mechanism's reliability in providing scores across various scales and demonstrating its capability in localizing scale-specific knowledge. The cross-scale attentions can be further utilized to offer AI-based guidance for knowledge exploration in pathological images in future work, distinguishing it from other methods.

\begin{table}[t]
\caption{Classification Performance on two toy dataset.}
\centering
\scriptsize
\setlength{\tabcolsep}{1mm}
\renewcommand\arraystretch{1}
\begin{tabular}{l|cc|cc|c}
\toprule
\multirow{1}{0.8in}{Method} & \multicolumn{2}{c}{Micro-anomaly}  & \multicolumn{2}{c}{Macro-anomaly} & \multirow{1}{0.3in}{Mean}\\
\cmidrule(lr){2-3}
\cmidrule(lr){4-5}
& AUC & AP & AUC & AP\\
\midrule
DeepAttnMISL(20$\times$)~\citep{yao2020whole}  &  1.0000 & 1.0000 &  0.7374& 0.9052 & 0.9107\\
DeepAttnMISL(10$\times$)~\citep{yao2020whole}  &  0.4898 & 0.5358 &  1.0000 & 1.0000 & 0.7564\\
DeepAttnMISL(5$\times$)~\citep{yao2020whole}  &  0.5408 & 0.5844 & 0.9992& 0.9998 & 0.7811\\
\midrule
Gated Attention~\citep{ilse2018attention}  & 0.7082 & 0.8082 & 0.6625& 0.8706 & 0.7624\\
DeepAttnMISL~\citep{yao2020whole} &  0.6879 & 0.7906 &  0.6547& 0.8143 & 0.7369\\
MS-DA-MIL~\citep{Hashimoto_2020_CVPR} & 1.0000 &1.0000 &   0.7468& 0.9085 & 0.9138 \\
DS-MIL~\citep{Li_2021_CVPR} & 0.4864& 0.5370 &   0.9919& 0.9977 & 0.7533\\
DTFD-MIL~\citep{zhang2022dtfd} &  0.4771& 0.5326 &  0.7563& 0.9223 & 0.7533\\
HAG-MIL~\citep{xiong2023diagnose} &  0.6483 & 0.7238 & 0.6979  & 0.8313 & 0.7343\\
\midrule
CS-MIL(Mean-vector) & 0.4429 & 0.5084 &  0.9893
& 0.1041 & 0.5112 \\
CS-MIL(Concatenation) & 0.4638 & 0.5316 &  1.0000
& 1.0000 & 0.7538 \\
CS-MIL(Ours) & 0.9627 & 0.9820 & 0.9982 &0.9994 & \textbf{0.9856}\\
\bottomrule
\end{tabular}
\label{tab:ablation1}
\end{table}

\section{Discussion}
Table~\ref{tab:Classification} and Fig.~\ref{fig3:Results} demonstrate that the multi-scale models performed better than the single-scale models, suggesting the usefulness of external information from multi-scale data on WSIs. The proposed CS-MIL model outperformed the other models in most evaluation metrics, highlighting the effectiveness of cross-scale attention, which holistically learns information from multiple scales and considers the cross-scale relationships in MIL.

Figure~\ref{fig5:ToydatasetResults} demonstrates that the CS-MIL model locates positive regions using instance scores, while the cross-scale attention maps identifies the correct scale where distinctive patterns occur. In Macro-anomaly dataset, the regions with larger circles are highlighted more at 5$\times$, providing further evidence that the model differentiate between shape patterns of ellipses and circles with larger visual fields.

To further investigate the efficacy of the cross-scale attention mechanism, we conducted experiments integrating cross-scale features through mean-vector and concatenation designs as maintaining a 1:1:1 ratio of attention scores. The performance was then evaluated on two real pathological datasets, showcasing the cross-scale attention mechanism's capability for pathological image classification. These designs were also applied to the two toy datasets, each embodying distinctive morphological patterns observed in digital pathology. In Table~\ref{tab:ablation1}, the performance of single-scale models indicates that only the micro white cross pattern could be captured at 20$\times$, while the macro ellipse and circle were differentiated across three scales. When analyzing the micro-anomaly dataset, where the pattern is present only at a single scale, refinement strategies (MS-DA-MIL, etc.) performed well in capturing target features. Conversely, concatenation strategies (DS-MIL, mean-vector, concatenation, etc.) were more effective in aggregating patterns across scales, resulting in superior performance on the macro-anomaly dataset. Distillation-based (hierarchical-based) methods (DTFD-MIL, HAG-MIL, etc.) achieve better performance when features are inheritable and region-consistent across hierarchical scales in the Macro-anomaly dataset, while failing to recognize scale-specific patterns lacking consistency across different scales in the Micro-anomaly dataset. The results demonstrate that the proposed cross-scale attention mechanism is both efficient and flexible, enabling improved pattern localization for scale-specific knowledge and enhanced pattern integration for hierarchical knowledge across different scales. These findings underscore the versatility of our proposed cross-scale attention mechanism in addressing different morphological patterns in digital pathology.

The parameter numbers and GPU memory usages of the state-of-the-art models in different multi-scale schemas are reported in Table~\ref{tab:ablation4}. The proposed method achieved better classification performance by only adding lower computational complexity.

\begin{table}
\centering
\scriptsize
\caption{Model Complexity}
\begin{tabular}{l|cc}
\toprule
Method & Parameters & GPU Mem\\
\midrule
DeepAttnMISL(5$\times$)~\citep{yao2020whole} & 135,362 & 1513 MiB\\
\midrule
MS-DA-MIL~\citep{Hashimoto_2020_CVPR} & 135,362 & 1601 MiB\\
HAG-MIL~\citep{xiong2023diagnose} & 406,086 & 1397 MiB\\
\midrule
CS-MIL(Ours) & 137,475 & 1493 MiB\\

\bottomrule
\end{tabular}
\label{tab:ablation4}
\end{table}

There are certain limitations and scope for improvement in our study. In the present model, the pretraining process is executed separately for three models at different scales, which requires significant computational resources and does not to capture inter-scale knowledge during self-supervised learning. An Omni model trained with images from multiple scales and imbued with scale-aware knowledge in the feature embedding is promising.

Moreover, the largest visual field of the current pipeline is 1024 $\times$ 1024 pixels, which is still a relatively small area in WSIs. However, the recent advancements in ViTs present an opportunity to enhance the pipeline by incorporating larger spatial relationships and more regional information in larger visual fields, allowing it to receive all information at the slide level directly.

Although the primary purpose of feature clustering for bag generalization is to aggregate patches that are sparsely distributed on WSIs without spatial constraints, it remains intriguing to conduct similarity analysis on features across different scales. This analysis aims to comprehensively understand the relationships between features across scales and the knowledge scenarios in real datasets, in comparison to the proposed toy dataset design.

\section{Conclusion}

In this study, we introduce a novel cross-scale MIL approach that effectively integrates multi-scale features with inter-scale knowledge. Additionally, the proposed method utilizes cross-scale attention scores to generate importance maps, enhancing the interpretability and comprehensibility of the CS-MIL model. Our experimental and simulated results reveal that the proposed approach outperforms existing multi-scale MIL benchmarks. The visualization of cross-scale attention produces scale-specific importance maps that potentially assists clinicians in interpreting image-based disease phenotypes. This contribution highlights the potential of cross-scale MIL in digital pathology and encourages further research in this area.

\section{Acknowledgements}
This work is supported by The Leona M. and Harry B. Helmsley Charitable Trust grant G-1903-03793, NSF CAREER 1452485, and Veterans Affairs Merit Review grants I01BX004366 and I01CX002171, and R01DK103831. This work is also supported in part by NIH R01DK135597 (Huo).

\bibliographystyle{model2-names.bst}\biboptions{authoryear}
\bibliography{refs}

\appendix

\newpage
 \renewcommand{\thesection}{\Alph{section}}

\end{document}